\documentclass[sigconf, review]{acmart}
\AtBeginDocument{%
  }

\usepackage{algorithm}
\usepackage{algorithmic}
\usepackage{booktabs}    
\usepackage{tabularx}    
\usepackage{array}       
\usepackage{booktabs}
\usepackage{array} 
\setcopyright{none}
\renewcommand\footnotetextcopyrightpermission[1]{}
\acmConference{JAUNT}{JAUNT}{2025}

\begin{document}

\title{JAUNT: Joint Alignment of User Intent and Network State for QoE-centric LLM Tool Routing}


\author{Enhan Li}
\email{lienhan_eee@connect.hku.hk}
\affiliation{
  \institution{The University of Hong Kong}
  \country{}
}

\author{Hongyang Du}
\email{duhy@hku.hk}
\affiliation{
  \institution{The University of Hong Kong}
  \country{}
}



\begin{abstract}
Large Language Models (LLMs) increasingly rely on emerging protocols such as the Model Context Protocol (MCP) to invoke external tools and services. However, current tool routing mechanisms remain fragile because they only consider functional matching between users' queries and tools. 
In practice, user intent expressed through queries can be vague or underspecified, and the actual Quality of Experience (QoE) also depends on external factors such as link latency and server availability that are not captured by semantics alone. 
To address this challenge, we propose JAUNT, a framework for {\underline{J}}oint {\underline{A}}lignment of {\underline{U}}ser intent and {\underline{N}}etwork state in QoE-centric {\underline{T}}ool routing.
JAUNT introduces a dual-view alignment strategy that interprets user intent while employing LLM agents to construct network profiles, mapping numerical performance indicators into the semantic space to guide routing. 
We further design a benchmark that integrates diverse user request patterns with heterogeneous network states, enabling systematic evaluation of QoE outcomes. 
Experimental results show that JAUNT significantly improves QoE compared with several baselines, demonstrating the importance of aligning both intent and network state for scalable LLM service orchestration.
\end{abstract}




\makeatletter
\let\@copyrightspace\relax
\makeatother
\pagestyle{empty}
\maketitle

\section{Introduction}
Large Language Models (LLMs) have evolved from text-based conversational systems~\cite{grattafiori2024llama3,gpt4,qwen3} into general-purpose intelligence engines interacting with external tools and services~\cite{zhuang2023toolqa,nam2024using,chang2024survey,ferrag2025llm,li2025review}. This transition reflects an inevitable trajectory toward integration, where LLMs no longer operate in isolation but act as coordinators across heterogeneous digital resources. Emerging protocols such as the Model Context Protocol (MCP)~\cite{mcp_introduction,mcp_landsacape,mcp_survey} have enabled LLMs to connect seamlessly with applications such as booking, design, learning, and data analysis within a unified conversational interface. Such integration transforms LLMs into a new operating paradigm that manages user intents, orchestrates tools, and delivers end-to-end task execution through natural dialogue~\cite{yang2023gpt4tools}. As this ecosystem expands, the ability of LLMs to understand user goals and invoke the right external services becomes a fundamental capability shaping the next generation of AI-driven platforms, analogous to an intelligent operating system for human–AI interaction~\cite{shah2025using,bodonhelyi2024user}.
 
While external tool integration greatly expands the functional capacity of LLMs, it also reveals the significant role of the network as both an enabler and a constraint of intelligence. 
Modern network infrastructure enables LLMs running on a host device to access diverse remote services through distributed computation, allowing a single model to compose tools across cloud platforms, knowledge bases, and specialized APIs in real time and thus form an open and scalable ecosystem unattainable by standalone models.
However, this advantage comes with inherent dependence on network conditions. Each tool invocation must traverse communication links and remote servers whose latency, bandwidth, and congestion vary with physical distance and traffic load. 
These variations influence the task response speed and the stability of task execution. 
For example, routing a user request to a distant or heavily loaded server may introduce substantial delay or timeout, degrading user-perceived Quality of Experience (QoE)~\cite{jain2004quality} despite the correct semantic match. 
However, current orchestration protocols, such as MCP, still treat the network as an ideal and static medium, overlooking its dynamic impact on service quality.

\begin{figure}[t]
\centering
\includegraphics[width=0.35\textwidth]{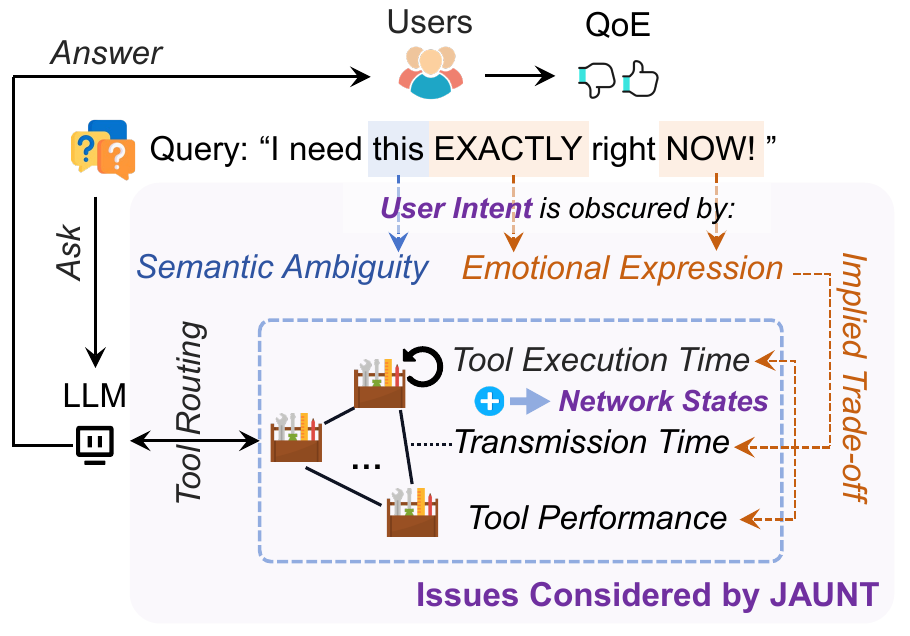}
\vspace{-0.2cm}
\caption{Motivation of JAUNT: Intent is obscured by semantic ambiguity, emotional expression, and network trade-offs.}
\label{motivationfigure}
\end{figure}
Incorporating network awareness addresses part of the orchestration challenge but does not fully determine the QoE.
The fundamental purpose of LLM-based systems is to serve diverse users, yet current routing methods often assume that user queries are clear and complete. In practice, this assumption rarely holds~\cite{shah2025using}. Public-facing LLM systems receive inputs that vary widely in specificity, tone, and contextual precision, leading to semantic ambiguity followed by inaccurate tool selection~\cite{wang2025intent}. More importantly, users' emotional expression in their queries inherently encodes expectations about network responsiveness and computational effort. 
Different user personalities imply distinct tolerances for latency, accuracy, and interaction cost~\cite{du2023attention}. For instance, a frustrated or time-sensitive user implicitly values low delay and immediate feedback, while an accuracy-focused user may prefer detailed responses even at longer execution times. 
As shown in Fig.~\ref{motivationfigure}, these latent expectations link user intent with network states, making them inseparable when maximizing users' QoE. 
However, current orchestration frameworks treat semantic understanding and network execution as independent processes, overlooking how emotional tone and behavioral patterns shape sensitivity to service performance. Bridging this gap requires a unified perspective that aligns user intent with dynamic network states to achieve effective and adaptive tool routing.

Addressing the above research gap requires tackling two challenges. The first is the lack of datasets that jointly reflect user intent diversity and network variability. Existing benchmarks mainly evaluate tool routing under well-defined queries and stable connections~\cite{guo2025mcp,gao2025mcp,wang2025mcp,fei2025mcp}, and while some works assess tool routing performance across diverse network conditions~\cite{li2025netmcp}, they fail to capture how real users express vague or emotionally biased requests that implicitly affect routing outcomes. 
The second challenge lies in LLM tool routing algorithmic integration, where semantic analysis of the user query must be aligned with network states to optimize end-to-end QoE.
To address these challenges, we first construct the Tool-Routing Intent and Persona (TRIP) benchmark for modeling user diversity.
TRIP includes user profiles capturing behavioral traits and queries generated from these profiles across common application scenarios such as travel planning and information retrieval. Each query is augmented with linguistic fuzziness and emotional tone to reflect realistic user intent and expression variations.
Building upon this, we leverage the intrinsic understanding capability of LLMs to jointly interpret users' long-term profiles, short-term emotional queries, and real-time network states to construct JAUNT, a framework for achieving {\underline{J}}oint {\underline{A}}lignment of {\underline{U}}ser intent and {\underline{N}}etwork state in QoE-centric {\underline{T}}ool routing. Our contributions are summarized as
\begin{itemize}
    \item We construct the TRIP benchmark to fill the current gap in evaluating LLM-based tool routing under diverse user intents. 
    TRIP provides structured user profiles and intent-driven query sets that capture realistic variations in linguistic fuzziness, tone, and behavioral preferences, enabling systematic analysis of subjective QoE in tool orchestration.
    \item We propose the JAUNT framework, which achieves dual-view alignment between semantic intent and network performance. An LLM agent infers both long-term intent from user profiles and short-term intent from current queries, then combines this with real-time network information to rank candidate tools and select the one expected to maximize user QoE.
    \item We conduct extensive experiments demonstrating that JAUNT balances task success rate, latency, and reliability to achieve high QoE, validating the importance of jointly aligning user intent and network state for LLM tool routing.
\end{itemize}

\section{Related Work}
\subsection{LLM Tool Routing and Evaluation}
Integrating external tools has transformed LLMs from static chatbots into interactive agents capable of executing complex tasks. 
A central challenge is tool routing, which determines how LLMs select and manage external services.
Early approaches treated routing as a semantic retrieval problem~\cite{yakovlev2024toolken,Rag-mcp}, where the LLM selects the most relevant tool schema based on textual similarity or prompt-level reasoning. 
Recent frameworks have expanded this view by introducing structured orchestration standards such as the MCP~\cite{mcp_introduction}, which unifies tool invocation through a standardized interface. Building on this foundation, various studies have examined routing efficiency, planning, and tool utilization from different perspectives. MCP-Zero~\cite{fei2025mcp} enables proactive toolchain construction by allowing models to retrieve and assemble tools dynamically. MCP-Bench~\cite{wang2025mcp} and MCP-RADAR~\cite{gao2025mcp} evaluate LLM agents on multi-step tool coordination, resource efficiency, and trajectory-level reasoning. MCP-AgentBench~\cite{guo2025mcp} further emphasizes end-to-end task success across realistic tool-mediated environments. 
However, these benchmarks generally assume that user instructions are explicit and well-structured, overlooking how real users express vague, underspecified, or emotionally charged intents. 
Although recent work, such as RECAP~\cite{mitra2025recap}, has begun addressing this issue by rewriting user dialogues into concise goal representations, these studies remain focused on language-level intent clarity and overlook how tone, responsiveness, and user expectations influence perceived QoE. 
Moreover, such factors implicitly encode user expectations about network responsiveness and computational performance~\cite{ramasamy2024latency,atilgan2025balancing}. Time-sensitive users often prioritize faster tool execution, while accuracy-oriented users may accept higher latency for reliable outcomes. These trade-offs are significant to real-world systems but are not reflected in current evaluation frameworks.

\subsection{Quality of Experience}
QoE provides a user-centered measure linking perceived satisfaction with system performance. In tool-augmented LLM services, QoE depends on the correctness of invoked tool outputs and on how efficiently and reliably the routing and execution process aligns with user expectations. Traditional QoE research in communication networks and multimedia systems focuses on latency, throughput, and reliability~\cite{atilgan2025balancing,ramasamy2024latency}, but such metrics alone cannot capture the experience in LLM-based orchestration, where satisfaction also depends on how well the system interprets user intent and delivers contextually appropriate results. 
Recent studies have examined related aspects from different angles. The study in~\cite{bodonhelyi2024user} showed that unclear or misrecognized intents often reduce perceived quality even when generated results remain accurate, while the work in~\cite{shah2025using} demonstrated that real user goals are diverse and context-dependent, indicating that uniform orchestration strategies cannot satisfy all users. In tool-based environments, this diversity translates into different tolerances for delay, precision, and computational cost, which collectively determine perceived QoE. Further, the framework in~\cite{wang2025intent} treated user intent as a latent representation of preference trade-offs, a perspective analogous to tool-routing agents that must infer implicit expectations about responsiveness and accuracy. Psychophysical analysis in~\cite{du2023attention} also revealed that perceived quality follows non-linear sensitivity patterns described by the Weber–Fechner law~\cite{dehaene2003neural}, where slight latency increases near perceptual thresholds cause disproportionate drops in satisfaction. These findings suggest that user intent and system responsiveness are inherently coupled, and achieving satisfactory QoE requires a joint modeling paradigm that captures both user diversity and operational dynamics for adaptive, user-centric tool routing.

\section{User-Centric QoE Modeling}
\label{Section3QoE}
We first introduce a general user-centric QoE modeling that captures both subjective and objective factors for users interacting with LLM agents. 
Conventional service evaluation for LLM-enabled tool-assisted tasks is often modeled as a binary outcome variable $c \in {0,1}$.
Here, $c=1$ denotes that the LLM agent successfully fulfills the user request, meaning the query is routed correctly and the tool executes as expected, while $c=0$ indicates failure.
Thus, the basic QoE is then defined as
\begin{equation}
\text{QoE}_{\text{base}}(c)=
\begin{cases}
Q, & c=1,\\[2pt]
0, & c=0,
\end{cases}
\end{equation}
However, this basic form ignores two effects that matter in practice. First, it ignores the impact of user waiting time. The end-to-end latency determined by both network transmission and tool execution, i.e., $L=t_{\text{net}} + t_{\text{tool}}$, directly affects user QoE. 
A long waiting time reduces perceived quality even when the final task output is correct. More importantly, prolonged latency and unstable networks also increase the probability of tool failure, such as timeouts or server crashes. 
Second, it ignores user heterogeneity and personalities. Users do not value correctness and latency equally. For example, a user issuing urgent translation requests prioritizes response speed, while another user relying on tool execution for detailed calculations prioritizes correctness. 

To capture these effects, we refine the QoE definition by incorporating the Weber-Fechner law into latency perception~\cite{dehaene2003neural,du2023attention}. The Weber-Fechner law states that the perceived change in stimulus intensity is proportional to the relative change of the physical stimulus. 
Specifically, the QoE penalty from the latency, i.e., $D(L)$, could be expressed in differential form as~\cite{dehaene2003neural}
\begin{equation}
\frac{\mathrm{d}D}{\mathrm{d}L} = \frac{w_1}{L/L_{\text{th}} + 1},
\end{equation}
where $w_1>0$ is a waiting time sensitivity parameter and $L_{\text{th}}$ is a reference latency threshold determined by specific user tasks. 
The denominator includes an additional constant term to avoid singularity at $L=0$ and to reflect the gradual saturation of human sensitivity to delay.
When $L$ is small, minor increases are obviously perceived by users, whereas the marginal impact becomes weaker once $L$ is already much larger than $L_{\text{th}}$. 
By integrating the differential relationship with respect to latency $L$, the corresponding QoE degradation function is obtained as
\begin{equation}
D(L) = w_1 \ln\!\Big(1+\frac{L}{L_{\text{th}}}\Big),
\end{equation}
which ensures $D(0)=0$ and produces a sublinear growth consistent with human perception.

Therefore, the conditional QoE becomes
\begin{equation}
\text{QoE}(c,L)=
\begin{cases}
w_2 Q - D(L), & c=1,\\[4pt]
- D(L), & c=0,
\end{cases}
\end{equation}
where $w_2\ge 0$ represents the task satisfaction coefficient associated with successful task completion, reflecting the user's perceived QoE gain when the LLM system delivers a correct result. A higher value of $w_2$ indicates that the user places greater emphasis on task success relative to latency penalties. 
The success probability for making $c=1$ is modeled as a single effective term
\begin{equation}
p_s(L,x;q,u)=p_{\text{route}}(q,u)\cdot p_{\text{tool}}(L)\cdot p_{\text{net}}(L,x),
\end{equation}
which depends on latency $L$, current network state $x$, user's query $q$, and user profile $u$. Here, $p_{\text{route}}(q,u)$ is the probability that the query $q$ is routed to the correct tool, $p_{\text{tool}}(L)$ is the probability that the selected tool executes successfully, and $p_{\text{net}}(L,x)$ denotes the probability that the underlying network remains stable without disconnection, failure, or interruption under state $x$. Each component represents an independent source of uncertainty contributing to the overall success probability $p_s(L,x;q,u) \in[0,1]$.

This refined QoE model serves as the baseline for later evaluations. The parameters $w_1$ and $w_2$ differ across users. Real systems can estimate them through online A/B testing or adaptive calibration based on behavioral traces such as satisfaction ratings~\cite{jia2025towards}. In the following benchmark analyses, we assume that $w_1$ and $w_2$ are quantifiable for evaluation while remaining unknown to LLM tool routing algorithms.


\section{TRIP Benchmark Construction}
\label{sec:dataset_generation}
In this section, we introduce the systematic methodology for constructing TRIP as a comprehensive and realistic benchmark to evaluate the QoE of LLM services under diverse user query characteristics. 
The generation process consists of four sequential stages, each designed to introduce specific dimensions of user diversity and query complexity that reflect real-world interaction patterns. 

\subsection{User Profile Generation}
Building on the analysis in Section~\ref{Section3QoE}, user behavior is primarily governed by two sensitivity parameters, i.e., waiting time sensitivity ($w_1$) and task satisfaction ($w_2$), which jointly determine how users perceive latency and correctness in service responses. To simulate realistic behavioral diversity, we generate user profiles with distinct combinations of  $w_1$ and $w_2$ that represent different personality archetypes. 
Each profile thus serves as a labeled instance in the TRIP dataset, guiding subsequent intent and query generation. We define a comprehensive taxonomy of four major user categories and nine representative subtypes, ensuring coverage across a broad spectrum of behavioral traits. 
This taxonomy is produced through a constrained random assignment process to emulate the diversity of users encountered in real-world services, as summarized in {\textbf{Algorithm~\ref{alg:profile_generation}}}. As shown in Fig.~\ref{tripfigure}, our user taxonomy comprises four major categories with detailed sub-types:
\begin{figure}[t]
\centering
\includegraphics[width=0.47\textwidth]{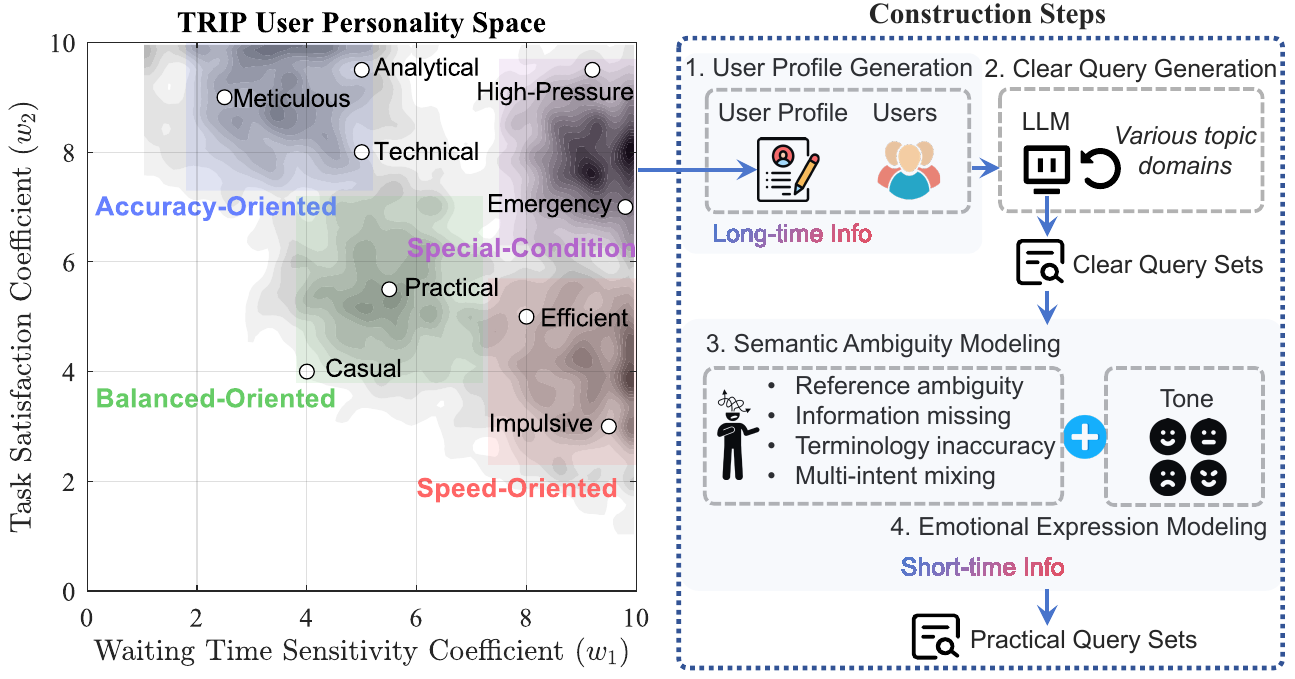}
\caption{Overview of TRIP benchmark construction pipeline.}
\label{tripfigure}
\end{figure}

\begin{algorithm}[t]
\caption{User Profile Generation Process}
\begin{algorithmic}[1]
\REQUIRE Number of users to generate: $n$, Topics list: $T$
\ENSURE Set of user profiles $U = \{u_1, u_2, ..., u_n\}$
\STATE Initialize empty set $U$, topic assignment counter $C$
\STATE Define 9 user types under 4 categories:
\STATE \quad $\mathcal{C}_{\text{speed}} \leftarrow \{\texttt{impulsive}, \texttt{efficient}\}$
\STATE \quad $\mathcal{C}_{\text{accuracy}} \leftarrow \{\texttt{meticulous}, \texttt{technical}, \texttt{analytical}\}$
\STATE \quad $\mathcal{C}_{\text{balanced}} \leftarrow \{\texttt{practical}, \texttt{casual}\}$
\STATE \quad $\mathcal{C}_{\text{special}} \leftarrow \{\texttt{emergency}, \texttt{high\_pressure}\}$

\FOR{$i = 1$ to $n$}
    \STATE $topic\_id \leftarrow (i-1) \mod |T|$ \COMMENT{Cyclic topic assignment}
    \IF{$i > |T|$}
        \STATE $topic\_id \leftarrow 0$ \COMMENT{Extra users assigned to first topic}
    \ENDIF
    
    \STATE Initialize available types $\mathcal{A} \leftarrow \mathcal{C}_{\text{speed}} \cup \mathcal{C}_{\text{accuracy}} \cup \mathcal{C}_{\text{balanced}} \cup \mathcal{C}_{\text{special}}$
    \STATE Remove types already assigned to current topic from $\mathcal{A}$
    
    \STATE Randomly select $user\_type$ from $\mathcal{A}$
    \STATE Assign $w_1, w_2$ based on predefined ranges for $user\_type$
    \STATE Generate descriptive profile using time tolerance guidelines
    \STATE Add user $u_i$ to set $U$
    \STATE Update topic assignment records
\ENDFOR
\RETURN $U$
\end{algorithmic}
\label{alg:profile_generation}
\end{algorithm}

\textit{Speed-First Category} prioritizes rapid response times over exhaustive accuracy:
\begin{itemize}
\item \texttt{Impulsive} ($w_1$: 9-10, $w_2$: 2-4): Makes rapid decisions with minimal verification, expects near-instant responses
\item \texttt{Efficient} ($w_1$: 7-9, $w_2$: 5-7): Values productivity and timely responses while maintaining basic accuracy standards
\end{itemize}

\textit{Accuracy-First Category} emphasizes thorough verification and comprehensive information:
\begin{itemize}
\item \texttt{Meticulous} ($w_1$: 1-4, $w_2$: 8-10): Requires thoroughly verified information regardless of time, can wait minutes for comprehensive results
\item \texttt{Technical} ($w_1$: 4-6, $w_2$: 7-9): Needs precise technical details but has reasonable time expectations
\item \texttt{Analytical} ($w_1$: 5-7, $w_2$: 9-10): Data-driven and requires detailed analysis with verification
\end{itemize}

\textit{Balanced Category} dynamically adjusts expectations based on situational context:
\begin{itemize}
\item \texttt{Practical} ($w_1$: 4-7, $w_2$: 4-7): Adapts expectations based on situation, balances speed and accuracy needs
\item \texttt{Casual} ($w_1$: 3-5, $w_2$: 3-5): Relaxed approach to both time and accuracy, values ease of use
\end{itemize}

\textit{Special Category} addresses high-stakes scenarios with extreme requirements:
\begin{itemize}
\item \texttt{Emergency} ($w_1$: 10, $w_2$: 6-8): Critical situations requiring both extreme speed and reasonable reliability
\item \texttt{High-pressure} ($w_1$: 9-10, $w_2$: 9-10): Demands both instant responses and perfect accuracy in high-stakes scenarios
\end{itemize}

Each generated user profile includes a unique identifier, its corresponding  ($w_1$ and $w_2$) coefficients, user type, topic assignment, and behavioral description. The algorithm guarantees logical consistency between user categories and sensitivity levels while maintaining balanced distribution across topics. This structure enables the TRIP dataset to model a realistic range of user personalities, providing a solid foundation for evaluating adaptive and user-centric tool routing strategies.

\subsection{Clear Query Generation}
For each generated user profile, multiple clear queries are auto-created by an LLM that reflect the user's likely information needs according to their behavioral characteristics. 
The generation process ensures that query content and structure are consistent with the user's profile, enabling realistic user-LLM agent interactions where response strategies must adapt to individual expectations. 
The queries span various topic domains such as travel planning, weather forecasting, restaurant recommendations, and technical information retrieval, which commonly require LLM agents with tool routing capabilities such as web search and file operations. 
Each query includes a user identifier, topic, and subtask complexity level aligned with the profile type. For example, time-sensitive users may raise tasks requiring rapid responses, and accuracy-oriented users may have queries demanding detailed and verified outputs.

\subsection{Semantic Ambiguity Modeling}
To replicate the natural vagueness and incompleteness often found in practical user queries, we introduce controlled semantic ambiguity into each generated query. 
This process enables the evaluation of how LLM-based tool routing algorithms handle imperfect and underspecified instructions. Four representative types of ambiguity are applied in the TRIP benchmark:
\begin{itemize}
    \item \textit{Reference ambiguity}: vague or implicit expressions (e.g., ``Check that site again'' instead of specifying the website).
    \item \textit{Information missing}: missing key details required for precise responses (e.g., ``Book me a flight'' without providing destination or date).
    \item \textit{Terminology inaccuracy}: inaccurate or colloquial usage of technical terms (e.g., ``computer memory is full'' instead of ``RAM usage is high'').
    \item \textit{Multi-intent mixing}: combining unrelated requests within a single query (e.g., ``Find good sushi places and email the menu to my friend'').
\end{itemize}
Each query retains its core intent to remain solvable while reflecting the imperfect communication patterns of practical user interactions.

\subsection{Emotional Expression Modeling}
Beyond semantic ambiguity, real users exhibit subjective expression styles shaped by their impatience, attention to detail, and emotional states. 
To capture this dimension, each query is infused with an emotional tone deterministically derived from the user's waiting time sensitivity $w_1$ and task satisfaction $w_2$.
Table~\ref{tab:tone_mapping} shows some examples of varying combinations of $w_1$ and $w_2$ that yield clearly differentiated emotional expressions, ranging from frustrated and angry users (high $w_1$, low $w_2$) to methodical and calm users (low $w_1$, high $w_2$). 
Each query is rewritten by an LLM to reflect the corresponding tone with an emotional intensity above $0.5$ on a normalized $[0,1]$ scale, ensuring salient yet realistic expression. 
This modeling introduces user-dependent subjective variation that affects the perceived trade-off between efficiency and accuracy, enabling the evaluation of QoE-driven routing behavior under realistic user patterns.


\begin{table*}[ht]
\centering
\caption{Mapping of User Profile Characteristics to Emotional Tone}
\vspace{-0.3cm}
\label{tab:tone_mapping}
{\small \begin{tabularx}{\textwidth}{@{}cccX@{}}
\toprule
\textbf{$w_1$} & \textbf{$w_2$} & \textbf{Tone Type} & \textbf{Example Expression} \\
\midrule
High & Low & Frustrated/Angry & ``Just find me some Italian restaurants already! Are you even trying? This is unacceptable! How hard can this be?'' \\
\addlinespace[1ex]
High & High & Demanding/Urgent & ``I need this EXACTLY right, NOW! Don't mess this up!'' \\
\addlinespace[1ex]
Low & High & Methodical/Obsessive & ``I need to know EVERY detail about Tokyo's weather—temperature, humidity, everything.'' \\
\addlinespace[1ex]
Low & Low & Casual/Indifferent & ``Meh, whatever Italian places are around I guess. Not that hungry anyway.'' \\
\bottomrule
\end{tabularx}}
\end{table*}

\section{JAUNT}\label{sec:JAUNT}

In this section, we present JAUNT, an LLM-driven tool routing framework that aligns user intent and network states to maximize QoE. 
The core principle of JAUNT is that effective routing requires not only understanding what the user asks for but also how the user implicitly expresses preferences regarding latency and accuracy. 
Therefore, the LLM agent in JAUNT resolves two layers of uncertainty when interpreting user requests. The first is \textit{semantic uncertainty}, where the explicit intent must be inferred from ambiguous query content. The second is \textit{subjective uncertainty}, where short-term emotional tone and long-term behavioral traits imply the user's desired balance between response quality and speed. 
To operationalize this design, JAUNT consists of three key modules:  
\begin{itemize}
    \item \textit{Semantic Intent Inference}: identifies the user's underlying goal and retrieves a top-$K$ candidate tool set through LLM-based semantic matching.  
    \item \textit{Network Latency Prediction}: estimates tool-side performance indicators, including latency and stability, to represent the real-time network state.  
    \item \textit{Joint QoE-centric Tool Routing}: employs an LLM agent to jointly reason over the inferred intent and predicted network state, selecting the tool that maximizes expected QoE.  
\end{itemize}
\begin{figure}[!t]
\centering
\includegraphics[width=0.47\textwidth]{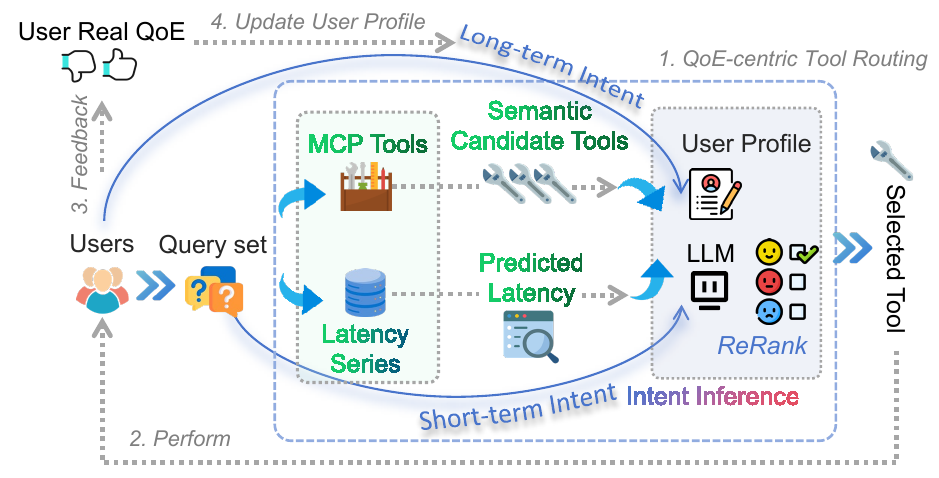}
\caption{Overall architecture of the JAUNT framework.}
\label{fig_4}
\end{figure}

\subsection{Semantic Intent Inference Module}\label{subsec:semantic_intent}
The semantic intent inference module aims to identify the most relevant candidate tools for a given user query through a two-stage retrieval process. This approach ensures both accuracy and computational efficiency in tool selection.

\textit{Stage 1: Tool Type Prediction using LLM.}
Given a user query $q$, we first employ an LLM to predict the tool type most likely required to fulfill the request. User queries often include redundant expressions and conversational fillers that vary with individual language habits, which may obscure the actual intent. The LLM extracts the essential functionality and produces a refined textual description $d_{\text{LLM}}$ as
\begin{align}
d_{\text{LLM}} = \text{LLM}_{\text{predict}}(q),
\end{align}
This preprocessing step preserves the core functional intent, providing a concise and semantically consistent representation for subsequent similarity matching and routing decisions.

\textit{Stage 2: Hierarchical Similarity Matching.}  
This stage performs hierarchical text matching to identify the most relevant tool candidates for the refined description $d_{\text{LLM}}$. We first compare $d_{\text{LLM}}$ with all MCP server descriptions $\mathcal{D}_{\text{servers}}$ to select the top-$K$ relevant servers as
\begin{equation}
\mathcal{S}_{\text{candidate}} = \text{Top}_k\big(\text{Sim}(d_{\text{LLM}}, \mathcal{D}_{\text{servers}})\big),
\end{equation}
where $\text{Sim}(\cdot)$ denotes a text similarity function such as BM25 or an embedding-based retrieval model.  
For each selected server $s_i \in \mathcal{S}_{\text{candidate}}$, the original user query $q$ is then matched against its internal tool descriptions $\mathcal{D}_{\text{tools}}^i$ to identify top-$M$ tool candidates:
\begin{equation}
\mathcal{T}_{\text{candidate}}^i = \text{Top}_m\big(\text{Sim}(q, \mathcal{D}_{\text{tools}}^i)\big).
\end{equation}
The final candidate set is formed by aggregating all tool candidates:
\begin{equation}
\mathcal{T}_{\text{candidate}} = \bigcup_{s_i \in \mathcal{S}_{\text{candidate}}} \mathcal{T}_{\text{candidate}}^i.
\end{equation}
This matching process progressively filters the search space, effectively balancing computational efficiency and semantic relevance.

\subsection{Network Latency Prediction Module}\label{subsec:latency_prediction}
The network latency prediction module predicts the end-to-end latency for each candidate tool, i.e., the user's waiting time, capturing both network transmission and tool execution delays. 
For each candidate tool, we employ the Exponentially Weighted Moving Average (EWMA) method~\cite{hunter1986exponentially} for latency prediction, which maintains a sliding window of recent latency measurements to adapt to temporal fluctuations:
\begin{align}
L_{\text{predicted}}(t_i) = \alpha \cdot L_{\text{current}}(t_i) + (1 - \alpha) \cdot L_{\text{historical}}(t_i),
\end{align}
where $\alpha$ is the smoothing factor, $L_{\text{current}}$ represents recent measurements, and $L_{\text{historical}}$ captures the aggregated long-term trend. 
By forecasting end-to-end delays, this module provides network state information about potential bottlenecks, including high latency or server congestion. 

\subsection{Joint QoE-centric Tool Routing Module}\label{subsec:joint_routing}
The joint QoE-centric tool routing module forms the core of the JAUNT framework. From a perception-decision perspective, effective tool routing depends on integrating multiple factors that influence the mapping between a user's raw query and the most appropriate tool.  

\begin{figure*}[!t]
\centering
\includegraphics[width=0.8\textwidth]{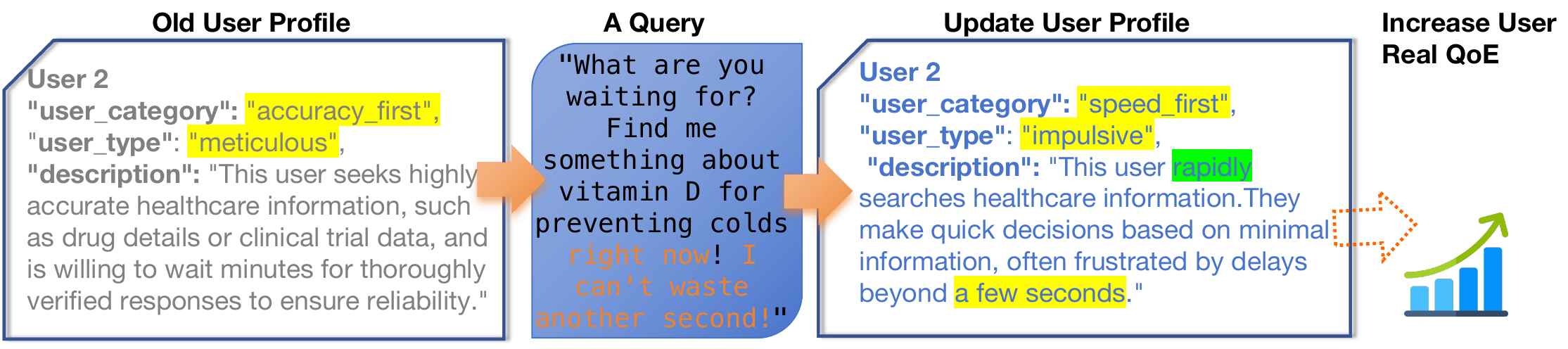}
\caption{User profile update mechanism. The profile is dynamically refined by extracting tone and implicit preferences from real-time queries, thereby enhancing the user's QoE.}
\label{fig:fig_user_profile}
\end{figure*}
From the perception side, effective tool routing depends on understanding two types of information within the user query. The first is the explicit functional demand, captured through the \textit{Semantic Intent Inference Module} as the scores of $\mathcal{T}_{\text{candidate}}$. The second is the implicit subjective tendency, expressed through user queries' tone, emotion, or phrasing, which reveals how the user values the trade-off between performance and latency. 
LLMs are particularly suited to perceive and interpret these cues, but additional contextual and system-level information is needed for making decisions. Specifically, the LLM agent maintains a \textit{User Profile} that describes long-term preferences learned from previous interactions, and obtains current network conditions from the \textit{Network Latency Prediction Module}. 
Thus, the LLM receives a structured input comprising four components:
\begin{enumerate}
\item \textit{Query Context}: The original user query $q$.
\item \textit{Semantic Similarity}: As introduced in Section~\ref{subsec:semantic_intent}, this represents the semantic matching score between the refined query $d_{\rm LLM}$ and each candidate tool's function description.
\item \textit{User Profile}: A structure maintained by JAUNT for each user over the long term, containing a description of the user's preferences regarding latency and task accuracy and a brief introduction to frequently used topics. It is periodically updated based on the user's query tone and QoE feedback, as shown in Fig.~\ref{fig:fig_user_profile}.
\item \textit{Predicted Latency}: As introduced in Section~\ref{subsec:latency_prediction}, this denotes the end-to-end latency estimated for each candidate tool based on the EWMA model.
\end{enumerate}

By jointly reasoning over these factors, the LLM agent selects tools to maximize QoE:
\begin{align}
t_{\text{selected}} = \arg\max_{t_i \in \mathcal{T}_{\text{candidate}}} \text{LLM}\left(
\begin{array}{l}
\text{user\_profile}(w_1, w_2), \\
\text{semantic\_sim}(q, t_i), \\
\text{predicted\_latency}(t_i), \\
\text{query\_context}
\end{array}
\right).
\end{align}
The routing decision jointly balances semantic relevance, user preference, and network performance. For time-sensitive users, JAUNT prioritizes tools with lower latency, while accuracy-oriented users are routed to those with higher functional matching. 
The LLM dynamically adapts short-term decisions based on the current query context, such as tone or urgency, while maintaining consistency with the long-term preferences recorded and updated in the \textit{User Profile}. Tools predicted to experience network congestion or latency spikes are proactively deprioritized, allowing the agent to select stable alternatives. Through this integrated reasoning, JAUNT achieves adaptive QoE-centric tool routing that aligns user intent with real-time network states.


\begin{figure}[t]
\centering
\includegraphics[width=0.45\textwidth]{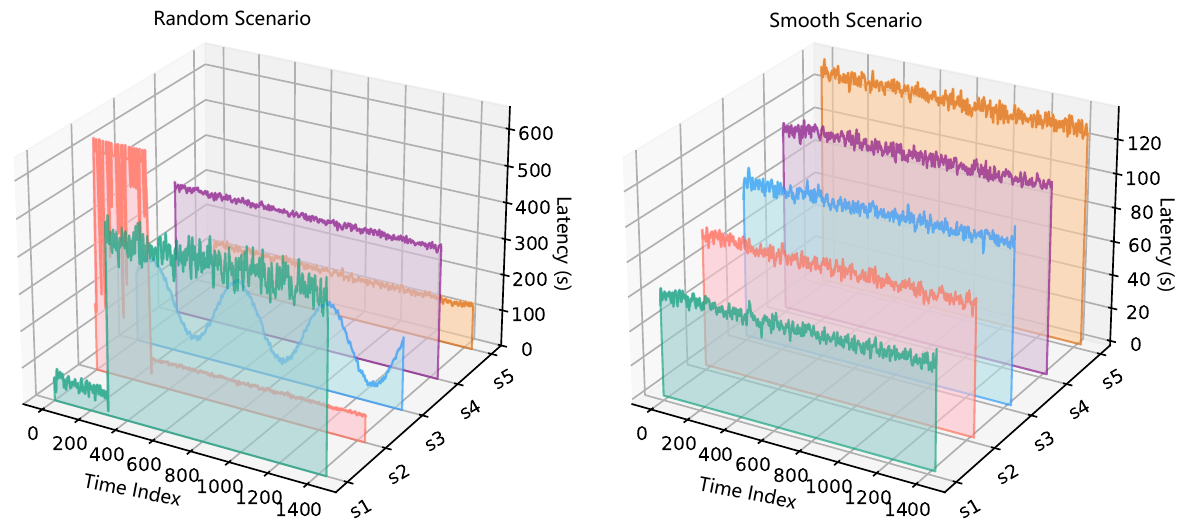}
\caption{Two network scenarios to evaluate JAUNT's performance under different conditions.}
\label{fig:latency}
\end{figure}

\begin{figure*}[t]
\centering
\includegraphics[width=0.9\textwidth]{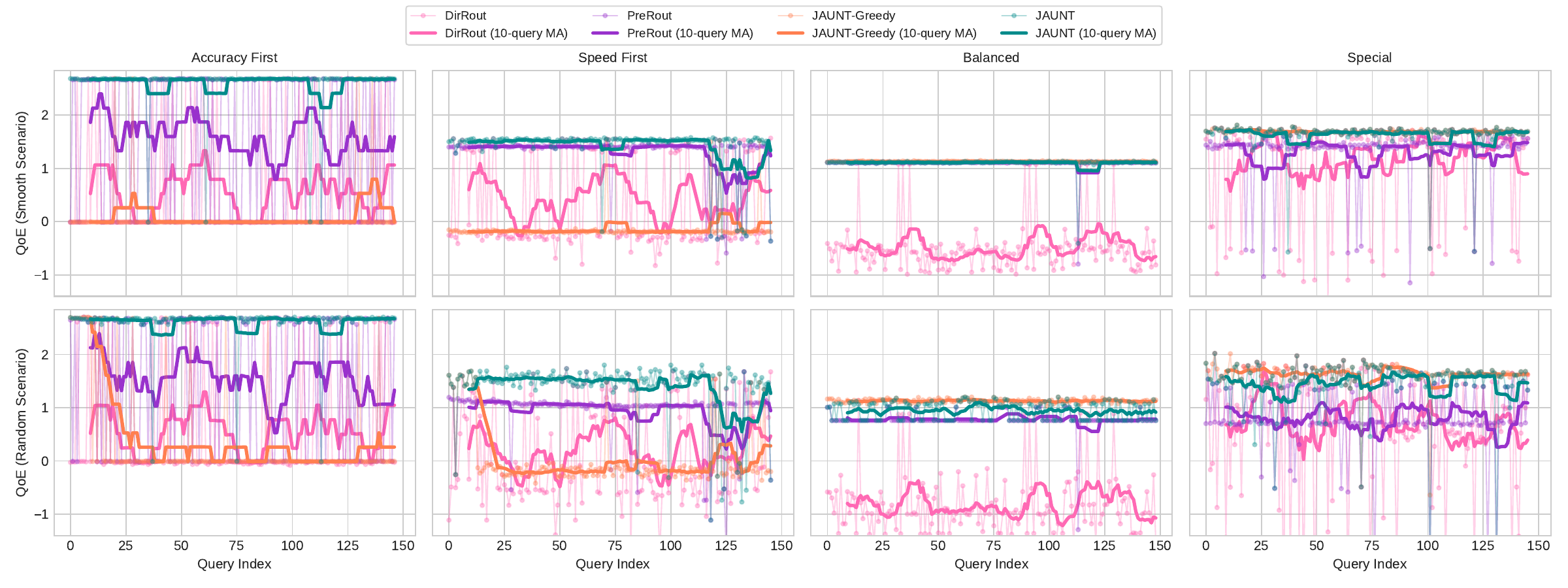}
\caption{Comparison of the average QoE across multiple queries for various routing algorithms under different network scenarios and user personalities. The 10-Query MA represents the average QoE over every 10 queries through moving average.}
\label{fig:qoe_metrics}
\end{figure*}

\begin{figure}[t]
\centering
\includegraphics[width=0.45\textwidth]{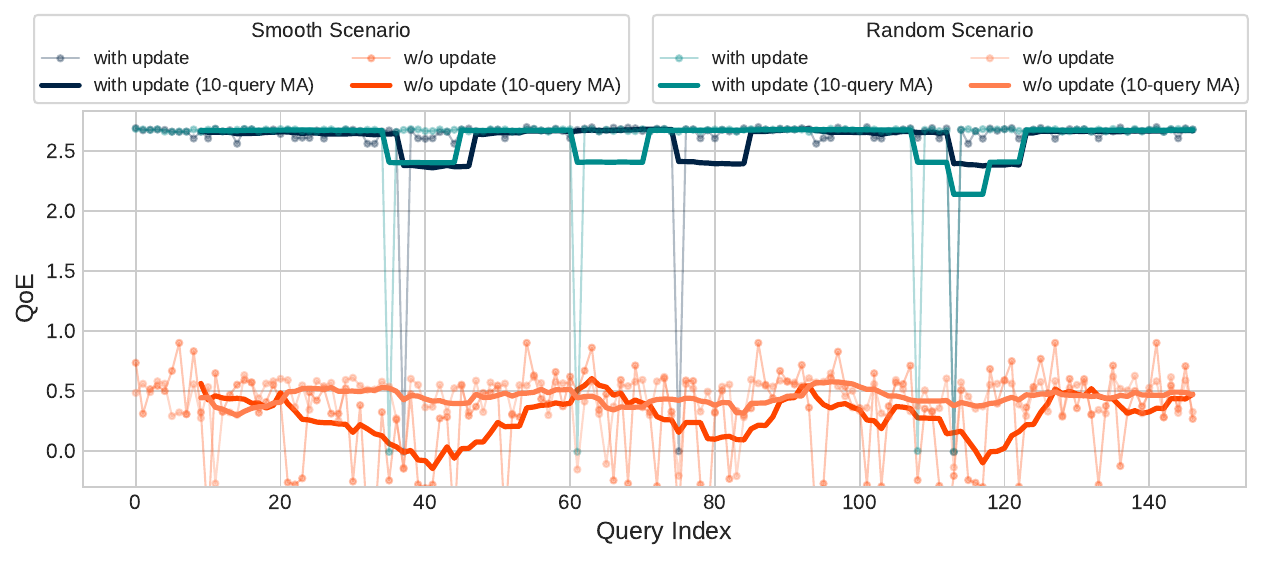}
\caption{Impact of long-term user profile update under different network scenarios. The 10-Query MA represents the average QoE over every 10 queries through moving average.}
\label{fig:qoe_updateprofile}
\end{figure}

\begin{figure*}[!t]
\centering
\includegraphics[width=0.9\textwidth]{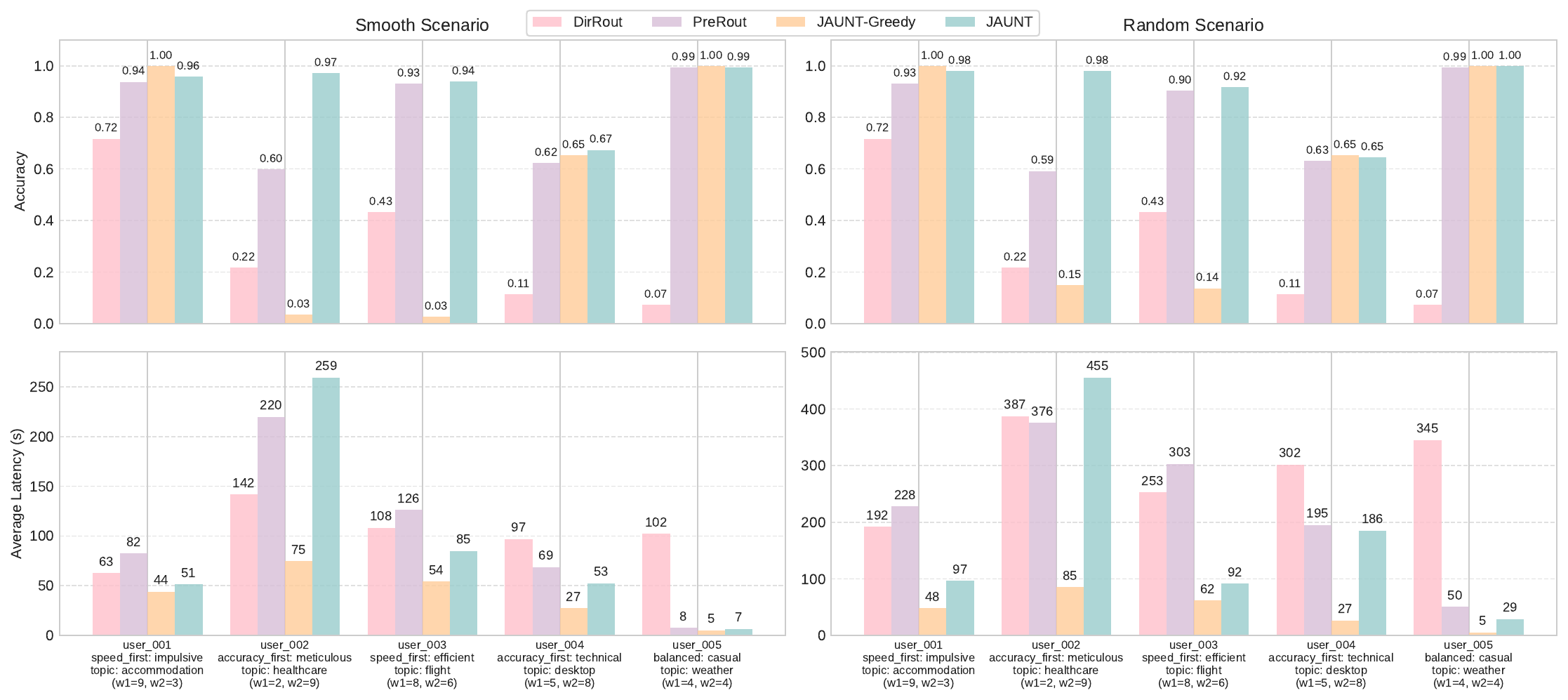}
\caption{Comparison of the average accuracy and latency across multiple queries for various routing algorithms under different network scenarios and user personalities.}
\label{fig:avg_metrics}
\end{figure*}

\section{Experiment}
\label{sec:Exp}
This section presents the experimental setup to validate JAUNT's performance. We construct a comprehensive test environment using the TRIP benchmark, comprising diverse topics, user profiles, and network conditions.

\subsection{Experiment Setup}

\textbf{MCP Server and User Configuration.}
NetMCP~\cite{li2025netmcp} provides a platform for simulating tool routing in real-world networks along with integrated tool routing algorithms. Based on NetMCP, we establish a comprehensive test environment comprising five distinct typical topics, including accommodation search, healthcare information, flight search, desktop operations and weather queries. We selected one real MCP server from remote mcp servers~\cite{smithery} for each topic and supplemented it with four mocked servers to create a robust testing set of 35 servers (7 real + 28 mocked). This configuration thoroughly evaluates JAUNT's routing capabilities across diverse functional domains while maintaining experimental control.
We generated 9 distinct user profiles using a structured prompt designed to capture realistic variations in user preferences. Each profile contains:
\begin{itemize}
\item \textit{Time Sensitivity ($w_1$)}: Ranging from 1 (very patient) to 10 (very impatient)
\item \textit{Accuracy Sensitivity ($w_2$)}: Ranging from 1 (content with quick answers) to 10 (requires detailed verification)
\item \textit{User Type}: Specific behavioral patterns (impulsive, meticulous, practical, etc.)
\item \textit{User Category}: High-level grouping (speed-first, accuracy-first, balanced, special)
\item \textit{Description}: Behavioral characteristics and implied time tolerance expectations
\end{itemize}

\textbf{Network Latency Simulation.}
We consider two network scenarios as shown in Fig.~\ref{fig:latency}. In the \textit{Smooth Scenario}, each topic was assigned a base latency value, with individual servers experiencing scaled variations through multiplicative factors and small variances, creating stable but differentiated latency curves. In the \textit{Random Scenario}, we implemented five distinct latency patterns across each topic's servers to simulate real-world network dynamics:
\begin{itemize}
\item \textit{Good-to-Bad Jitter}: Transition from low latency with rapid fluctuations to high latency with increasing variance
\item \textit{Bad-to-Good Stable}: Transition from high latency with intermittent peaks to low stable latency
\item \textit{Stable Fluctuating}: Consistent baseline with periodic wave-like oscillations
\item \textit{Stable High Latency}: Consistently elevated latency with minimal noise
\item \textit{Stable Normal Latency}: Consistently normal latency with minimal noise
\end{itemize}
These complex latency patterns enable a comprehensive and practical evaluation of different LLM tool routing algorithms' capabilities.

\textbf{Baselines.}
We implement three LLM tool routing algorithms as baselines for comparative analysis: 
\begin{itemize}
    \item \textit{Direct-Routing (DirRout)}: This baseline directly aligns user queries with the most relevant MCP tool using BM25-based semantic matching.
    \item \textit{Prediction-Routing (PreRout)}: Building upon DirRout, PreRout first employs an LLM to predict the most likely tool category for a given query and then performs semantic matching within that category to identify a ranked list of candidate tools.
    \item \textit{JAUNT-Greedy}: Built upon the JAUNT framework, JAUNT-Greedy follows the same processing pipeline but simplifies the {\textit{Joint QoE-centric Tool Routing Module}} to select the tool associated with the lowest latency directly.
\end{itemize}





\subsection{Results and Analysis}

{\textit{Effectiveness.}} Figure~\ref{fig:qoe_metrics} compares JAUNT with three baselines across four representative user types. Overall, JAUNT consistently achieves the highest or near-highest QoE in both smooth and random network scenarios, demonstrating its robustness across user preferences. The only exceptions appear in the \textit{Balanced} and \textit{Special} categories, where JAUNT-Greedy slightly surpasses JAUNT. This occurs because these two user types have comparable weights on task success and latency tolerance, resulting in a smaller performance gap between routing strategies. In such cases, JAUNT's joint reasoning process introduces marginally longer waiting times compared with the direct decision rule in JAUNT-Greedy, leading to a slightly lower perceived QoE. Nevertheless, JAUNT exhibits superior adaptability to dynamic network conditions and user heterogeneity. It maintains stable QoE trajectories across varying users and network scenarios, reflecting the advantage of its integrated LLM reasoning and temporal context modules. In contrast, JAUNT-Greedy shows larger fluctuations, particularly for user types with divergent preferences such as \textit{Accuracy First} and \textit{Speed First}, and its performance is more sensitive to network variations.

{\textit{Effect of Long-Term User Profile Update.}} Figure~\ref{fig:qoe_updateprofile} examines the influence of the long-term user profile module in JAUNT by comparing routing performance with and without user profile updates under both smooth and random network conditions. The user profile encodes historical behavior and preference patterns, enabling the routing algorithm to interpret user intent jointly with real-time network states and thus make more balanced decisions between task success and latency. As shown in the figure, maintaining up-to-date user profiles substantially improves QoE stability and average performance. When outdated profiles are used, the model's interpretation of user intent becomes inconsistent with the current context, leading to suboptimal routing choices and large QoE fluctuations, particularly under random network variations. In contrast, with profile updates (implemented as described in Fig.~4), JAUNT maintains a steady QoE trajectory across queries, demonstrating that adaptive user modeling enhances both responsiveness to network dynamics and alignment with user intent over time.

{\textit{Source of Performance Gain.}} Figure~\ref{fig:avg_metrics} analyzes the origin of JAUNT's performance advantages by comparing the average accuracy and latency across five users with distinct preference settings generated by the TRIP benchmark. Each user represents a different balance between waiting time sensitivity (i.e., $w_1$) and task satisfaction (i.e., $w_2$), allowing us to examine how routing strategies adapt to diverse personalities and network conditions. The results show that JAUNT consistently achieves the best trade-off between accuracy and latency, maintaining high QoE across all user types. In contrast, JAUNT-Greedy aggressively minimizes latency but often sacrifices task success. For example, for \textit{user\_002}, JAUNT-Greedy attains the lowest latency among all algorithms, but the accuracy drops to only $0.03$, indicating frequent routing failures caused by overly opportunistic tool selection. On the other hand, JAUNT strategically accepts slightly higher latency to preserve task correctness and overall QoE. This behavior reflects the benefit of incorporating LLM-based reasoning and the designed joint optimization module, which allows JAUNT to interpret user intent and network dynamics holistically rather than relying on single-factor heuristics. The consistent performance across both smooth and random network scenarios further demonstrates that JAUNT can generalize well across heterogeneous user demands and environmental uncertainties.

\section{Conclusion}\label{sec:Conclusion}
In this paper, we propose JAUNT, a framework for the joint alignment of User intent and network state in QoE-centric tool routing for LLM agents. We considered the limitations of current routing mechanisms that rely solely on semantic matching and ignore how network dynamics and user intent jointly determine user QoE. 
We constructed the TRIP benchmark to emulate realistic interactions by combining user profiles, emotional expressions, and network variations, enabling systematic evaluation under heterogeneous conditions. 
Experimental results showed that JAUNT consistently achieved higher QoE compared with several baselines. 
The framework effectively adapts to diverse user categories and varying network states, maintaining stable QoE trajectories through its dual-view reasoning and long-term user profile updating modules. These findings demonstrated that aligning intent understanding with network perception was essential for scalable and user-centric orchestration of LLM-driven tool ecosystems. Future work could extend JAUNT toward multi-agent routing and cross-platform coordination to support real-time and distributed LLM services.

\bibliographystyle{ACM-Reference-Format}
\bibliography{sample-base}

\end{document}